\documentclass[12pt]{iopart}
\usepackage{graphicx}
\usepackage{caption2}

\def\lsim{\mathrel{\raisebox{-.6ex}{$\stackrel{\textstyle<}{\sim}$}}}
\def\gsim{\mathrel{\raisebox{-.6ex}{$\stackrel{\textstyle>}{\sim}$}}}

\begin{document}

\title{Photon-pair jet production via gluon fusion at the LHC}

\author{Qiang Li}
\address{School of Physics, and State Key Laboratory of Nuclear Physics and Technology, Peking University, Beijing, 100871, People's Republic of China}

\author{Gao Xiangdong}
\address{Institute of High Energy Physics and Theoretical
Physics Center for Science Facilities, Chinese Academy of Sciences,
Beijing 100049, People's Republic of China}

\begin{abstract}
Photon-pair direct or jet-associated productions are important for relevant standard model measurement, Higgs and new physics searches at the LHC. The loop-induced gluon-fusion process $gg \to \gamma\gamma g$, which although formally  contributes only at the next-to-next-to-leading order to $\gamma\gamma j$ productions, may get enhanced by the large gluon flux at the LHC. We have checked and confirmed previous results on $gg\to \gamma\gamma$~\cite{dicus:1988}, $\gamma\gamma g$~\cite{deFlorian:1999tp} at one loop, using now the traditional Feynman diagram based approach and taking into account the quark mass effects, and further updated them for the 7 and 14 TeV LHC with new inputs and settings. We provide the details and results of the calculations, which involves manipulation of
rank-5 pentagon integrals. Our results show that the gluon-fusion process can contribute about $10\%$ of the Born result, especially at small $M_{\gamma\gamma}$ and $P^T_{\gamma\gamma}$, and increase further the overall scale uncertainty. Top quark loop effects are examined in detail, which shows importance near or above the threshold $M_{\gamma\gamma}\gsim 2m_t$.
\end{abstract}


\maketitle
\newpage

\section{Introduction}
\label{intr}
The Large Hadron Collider (LHC) is running smoothly with unprecedented high collision energy and luminosity which are necessary
for discovering Higgs particles or new physics beyond the Standard Model (SM). However, the higher the collision energy is, the more complex event topology gets involved, especially, hadron collision events with multi hard particles and large jet multiplicities become more probable, which deserve careful treatment.

Among various projects at the LHC, di-photon measurement is not only important for testing the Standard Model, but also crucial for understanding the background of Higgs search in low mass region, e.g., $m_h\lsim 140$\,GeV~\cite{haa}, so as for new physics search, e.g. extra dimension search in which higher invariant mass of di-photon gets concerned~\cite{Chatrchyan:2011jx}.

At the leading order (LO), di-photon production arises from $q\bar{q}$ annihilation process. The next-to-leading order (NLO) QCD correction was calculated decades ago~\cite{nloaa} and found to be very large, e.g. $\sigma^{\rm NLO}/\sigma^{\rm LO} \gsim 3$ for $M_{\gamma\gamma}\gsim 100\,$GeV at the 14 TeV LHC~\cite{Catani:2011qz}. The full next-next-to-leading order (NNLO) QCD prediction has been presented recently in Ref.~\cite{Catani:2011qz}, which gives additional 60\% enhancement at the LHC over the NLO one. Various pieces of NNLO calculations were presented before in Ref.~\cite{Barger:1989yd} for $\gamma\gamma+2j$ at LO, Ref.~\cite{Bern:1994fz} for $\gamma\gamma+j$ at one loop and Ref.~\cite{Anastasiou:2002zn} for $\gamma\gamma$ at two loop. The loop-induced gluon fusion (GF) contribution~\cite{dicus:1988}, appearing first at NNLO, was shown to be as sizable as LO result, especially for low di-photon invariant mass. The NLO corrections to the GF contribution has also been achieved in Ref.~\cite{Bern:2002jx} and found to be modest ($\lsim 10\%$ for $M_{\gamma\gamma}\gsim 100\,$GeV). Moreover, photons can also arise from fragmentation subprocesses of QCD partons which involves non-perturbative information on the parton fragmentation functions of the photon. One can suppress the fragmentation contributions by using the photon-jet isolation cut, e.g. the one proposed by Frixione~\cite{Frixione:1998jh}.

The NLO QCD and fragmentation contributions have been implemented into the fully differential tools, e.g. DOPHOX~\cite{Binoth:1999qq}, 2gammaMC~\cite{Bern:2002jx} and MCFM~\cite{Campbell:2011bn}, at parton level; Resbos~\cite{Balazs:2007hr} with transverse-momentum resummation; and POWHEG~\cite{D'Errico:2011sd} at hadron level interfacing with parton shower codes.

On the other hand, di-photon jet associated production, although more complex, might benefit from the the accompanying jet to increase the signal to background ratio by refining the experimental cuts~\cite{Mellado:2007fb}. The NLO corrections to $pp\to \gamma\gamma j$ was performed in Ref.~\cite{DelDuca:2003uz}, and found to be quite large, e.g. $\sigma^{\rm NLO}/\sigma^{\rm LO}$ varies from $3$ or $2$ for $M_{\gamma\gamma}$ ranging from 80 to 160 GeV at the 14 TeV LHC.

The gluonic NNLO finite subset, $gg\to \gamma\gamma g$, has been calculated in Refs.~\cite{deFlorian:1999tp,Balazs:1999yf} and found to be modest which is always smaller than $20\%$ of the Born result of $pp\to \gamma\gamma g$ for $M_{\gamma\gamma}>80\,$GeV. The calculations employed string-based methods~\cite{Bern:1993mq} and considered 5 massless flavors in the fermion loop.

In this work, we are revisiting and updating the calculations and results of
the GF process $gg\to\gamma\gamma, \, \gamma\gamma g$ at the 7 TeV and 14 TeV LHC~, with the traditional Feynman diagram based approach, which is straightforward to include the quark mass effects and top quark contributions. The paper is organized as follows. In Section~\ref{calculations} we describe the calculation. In Section~\ref{results} we present numerical results and discussions. Finally we conclude in Section~\ref{summary}.

\section{Calculation}
\label{calculations}
\begin{figure}[h]
\centering
\includegraphics[scale=0.8]{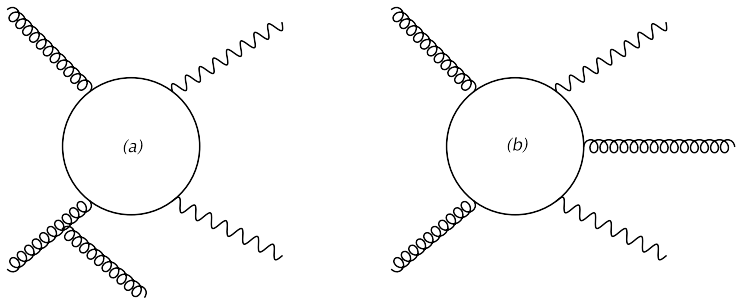}
\caption{Feynman Diagrams generated by Jaxodraw~\cite{Jaxodraw} for the partonic process $gg\rightarrow \gamma\gamma g$, sorted into 2 topological classes. Taking into account of all possible permutation, one gets 9 diagrams for (a) and 12 ones for (b). In addition, one needs to sum over the Fermion flavors and flow directions within the Fermion loop.}
\label{fd}
\end{figure}

The relevant one-loop Feynman diagrams and amplitudes for the partonic process $gg\rightarrow \gamma\gamma g$
have been generated with FeynArts 3.6~\cite{FeynArts}. The diagrams are sorted
into 2 topological classes, corresponding to boxes and pentagons,
as shown in Fig.~\ref{fd}.

The resulting Feynman amplitudes are then manipulated with FormCalc 6.2~\cite{FormCalc}. The Fortran libraries generated with FormCalc are linked with our phase space integration code to obtain the numerical results on total and differential cross sections. The tensor integrals are evaluated with the help of the LoopTools-2.5 package \cite{FormCalc}, where we have implemented the reduction method for pentagon tensor integrals up to rank 5 as proposed in Ref.~\cite{PentagonB}. We have also implemented in LoopTools the so called Alternative Passarino-Veltman Reduction for triangle and box tensor integrals~\cite{PentagonB} to improve numerical stability.

We have checked the cancellation of ultraviolet and infrared divergences in our calculations, and tested the independence on the mass scale in the Fermion loop. We have also further tested our working line by checking with previous results on $gg\to \gamma\gamma$~\cite{dicus:1988} and $\gamma\gamma g$~\cite{deFlorian:1999tp}, with agreement within uncertainty got by taking their inputs.

\section{Numerical Results}
\label{results}
In this section we present the total cross sections and differential
distributions for $\gamma\gamma$ and $\gamma\gamma+j$ productions at the LHC running with the collision energy at 7 TeV and 14 TeV, respectively. We impose the following set of cuts
\begin{eqnarray}\label{cut1}
&& |\eta_{j}|< 4.5 \,,\,\qquad P_T^{j}>50\,{\rm GeV}\,
\end{eqnarray}
to identify massless partons with jets. Photons are required to satisfy
\begin{eqnarray}\label{cut2}
&& |\eta_{\gamma}|< 2.5 \,,\,\quad P_T^{\gamma}>20\,{\rm GeV}
\,,\, \quad
R_{\gamma\gamma}= \sqrt{\Delta \eta^2+\Delta \phi^2}> 0.4\,.
\end{eqnarray}
Here $\eta$ is the
pseudorapidity and $\phi$ is the azimuthal angle around the
beam direction. Invariant mass of the photon-pair is further asked to satisfy
\begin{eqnarray}\label{cut3}
100\,{\rm GeV}<M_{\gamma\gamma}<700\,{\rm GeV}\,.
\end{eqnarray}
Moreover, we take the photon-jet isolation cut~\cite{Frixione:1998jh}
\begin{eqnarray}\label{cut4}
P_T^{j} \leq P_T^{\gamma}\frac{1-\cos R_{\gamma\, j}}{1-\cos \delta_0},\quad
{\rm for}  \quad R_{\gamma\, j}<\delta_0 = 0.7,
\end{eqnarray}
which can suppress fragmentation contributions efficiently.

Throughout our calculations, we set the electromagnetic coupling to $\alpha_{em}=1/132.5$ and top quark mass to $m_t=173.0$\,GeV, and employ the CTEQ6L1 parton distribution functions \cite{Pumplin:2002vw}
with the default LHAPDF~\cite{Whalley:2005nh} strong coupling value $\alpha_s(M_Z)=0.129783$.  Our
default choice for the renormalization and factorization scales is
$\mu_0=\sqrt{M^2_{\gamma\gamma}+P_{T\,\gamma\gamma}^2}$.

\begin{figure}[h]
\centering
\includegraphics[width=0.8\textwidth]{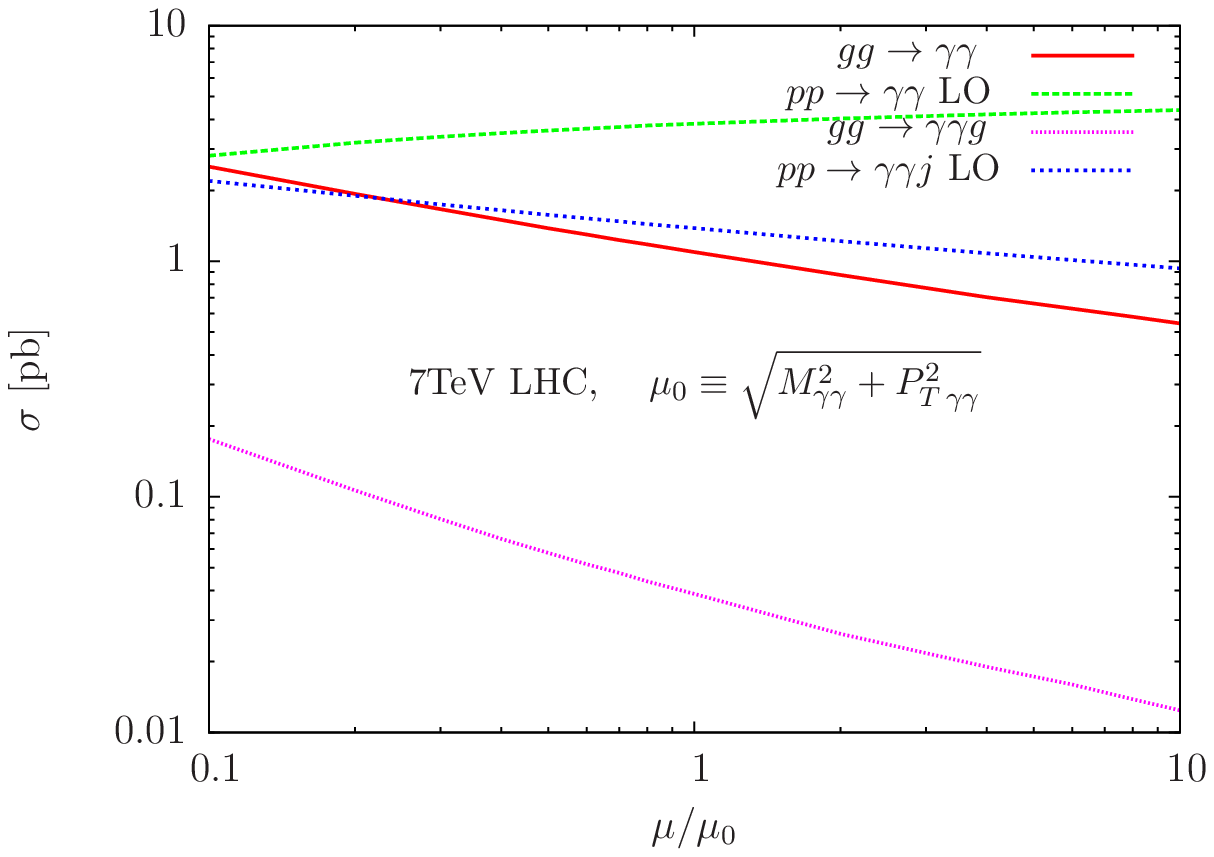}\vspace{0.1cm}
\includegraphics[width=0.8\textwidth]{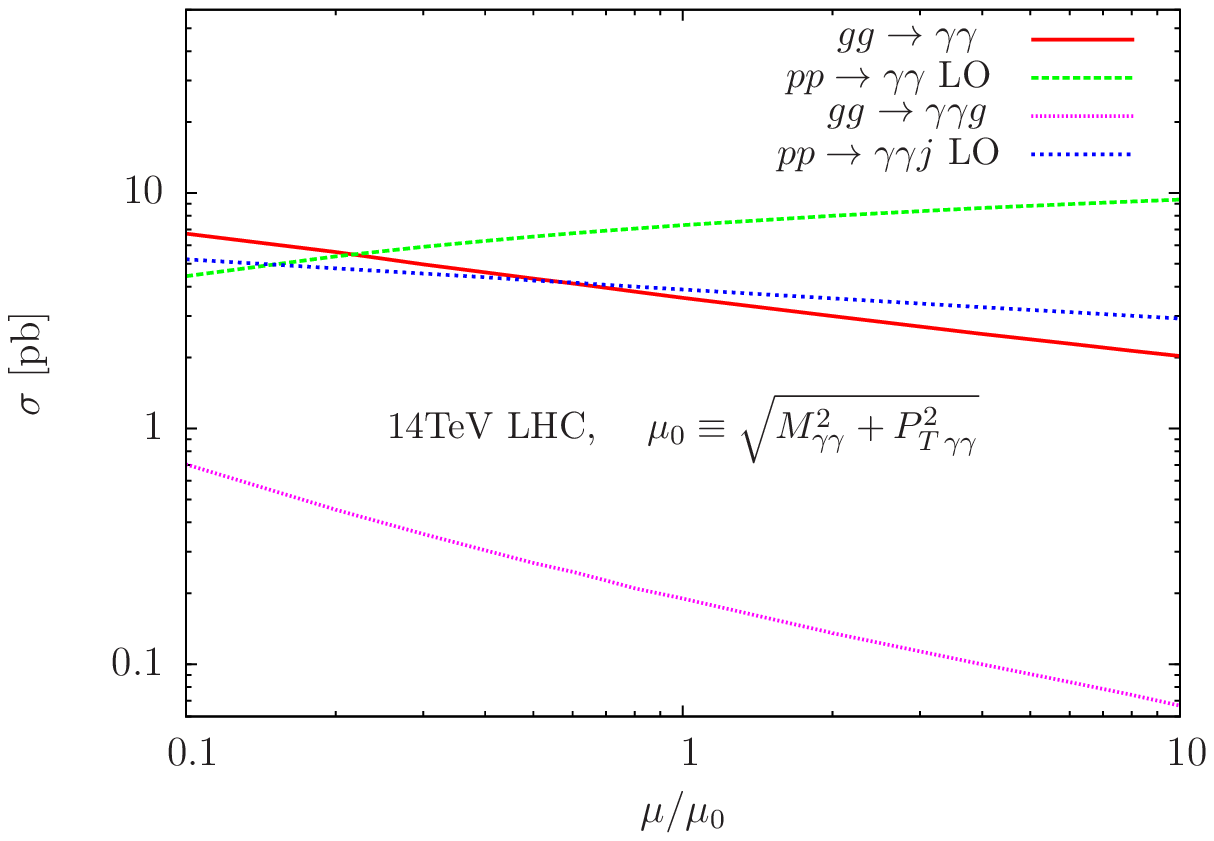}
\caption{Scale dependence of the GF and LO QCD total cross sections for
$\gamma\gamma$ and $\gamma\gamma j$ productions at the 7 and 14 TeV LHC, with $\mu_r=\mu_f=\mu$.}
\label{scale}
\end{figure}

Fig.~\ref{scale} shows dependences on the renormalization and factorization scales ($\mu_r=\mu_f=\mu$) of the GF and LO QCD $\gamma\gamma$ and $\gamma\gamma j$ production rates at the 7 and 14 TeV LHC.
The LO curves are got with the help of MadGraph4~\cite{Alwall:2007st}. One can see that the GF production rates
are significant (tiny) at small (large) scale, as expected due to gluon flux. At the 7 TeV LHC, the GF result counts 90\% and 12.4\% of the LO ones for $\gamma\gamma$ production~\footnote{Note from Ref.~\cite{Catani:2011qz}, the NLO QCD K factor is about 4 to 5 and the NLO cross section is quite stable, so we estimate the ratio of GF over NLO total cross section to be about 20-30\% at $\mu=0.1\mu_0$.}, while 8\% and 1.3\% for $\gamma\gamma j$ production, for $\mu=0.1\mu_0$ and $\mu=10\mu_0$, respectively. At the 14 TeV LHC, the GF result counts 151\% and 22\% of the LO ones for $\gamma\gamma$ production, while 13.5\% and 2.3\% for $\gamma\gamma j$ production, for $\mu=0.1\mu_0$ and $\mu=10\mu_0$, respectively. The GF results also enlarge the scale uncertainty for $\gamma\gamma j$: varying $\mu$ from $0.1\mu_0$ to $10\mu_0$, one gets uncertainties of $\pm 28.2\%$ and $\pm 33.0\%$, for the LO and LO+GF results at the 14 TeV LHC, respectively.

\begin{figure}[h]
\centering
\includegraphics[width=0.9\textwidth]{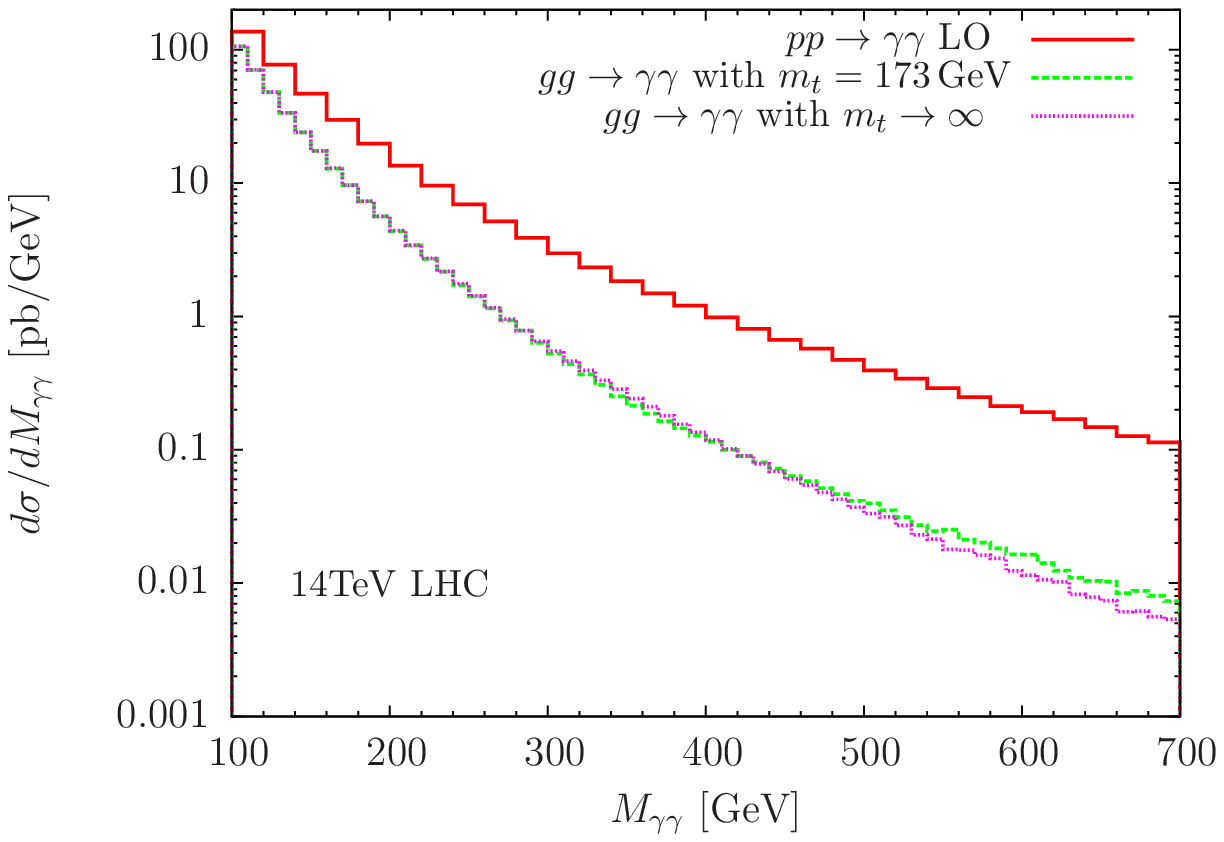}\vspace{0.1cm}
\includegraphics[width=0.9\textwidth]{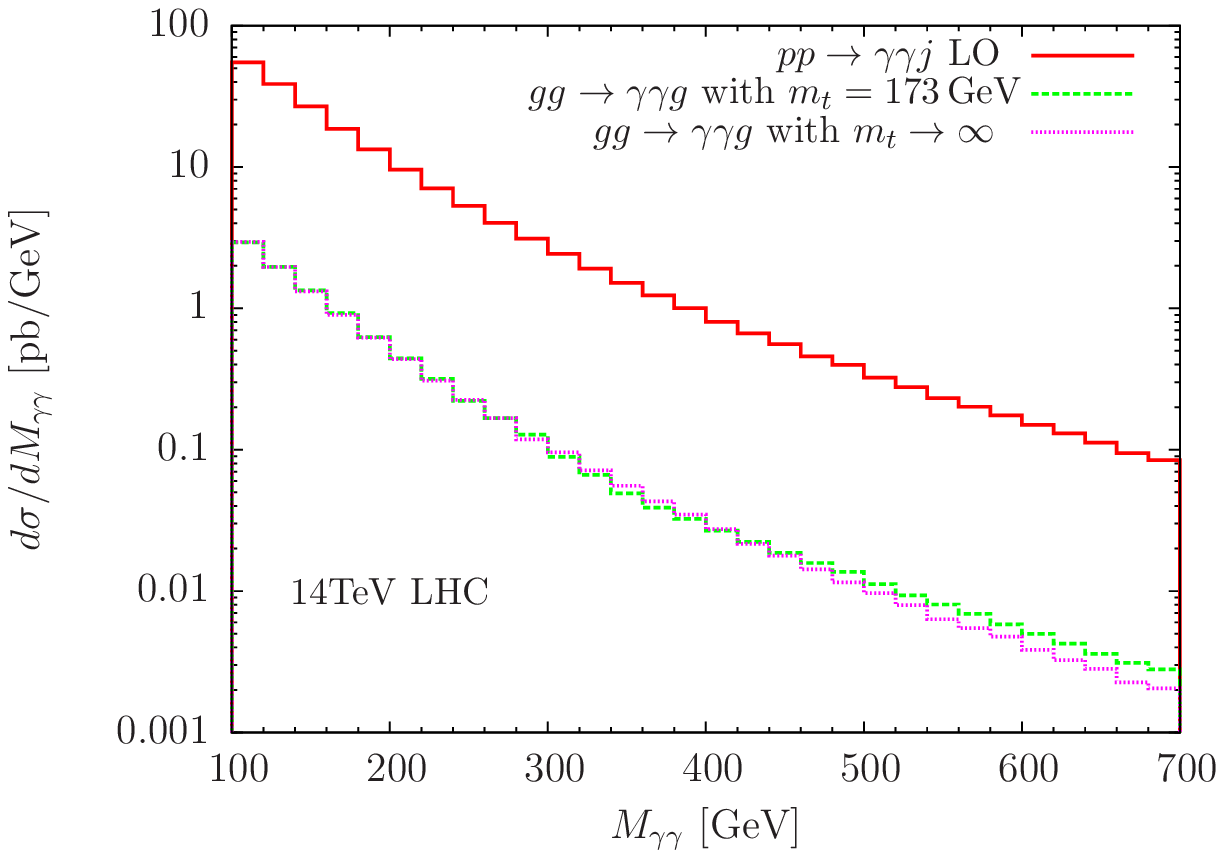}
\caption{Di-photon invariant mass distributions of the GF and LO QCD total results for $\gamma\gamma$ and $\gamma\gamma j$ productions at the 14 TeV LHC. The GF results w/o top loop are also shown.}
\label{maa}
\end{figure}

\begin{figure}[h]
\centering
\includegraphics[width=0.8\textwidth]{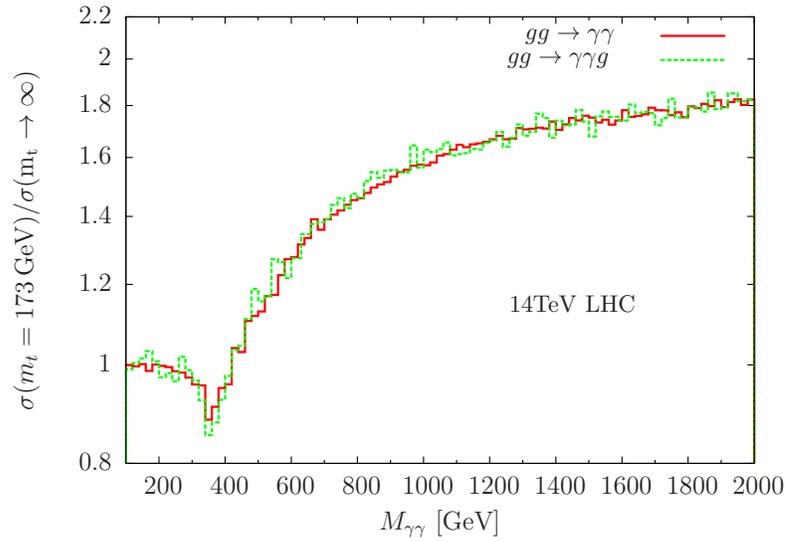}
\caption{Top quark mass effects on the production rates of $gg\to\gamma\gamma$ and $gg\to\gamma\gamma g$.}
\label{rat}
\end{figure}

Fig.~\ref{maa} displays the distributions of di-photon invariant mass $M_{\gamma\gamma}$, for both $\gamma\gamma$ and $\gamma\gamma j$ productions via GF at the 14 TeV LHC, in comparison with the LO QCD results. One can see that the GF curves tend to be softer than the LO ones, especially in $\gamma\gamma$ case. At $M_{\gamma\gamma}\sim 100\,$GeV, the GF counts about 78\% and 5.4\% of the LO results, for $\gamma\gamma$ and
$\gamma\gamma j$ productions, respectively.  At $M_{\gamma\gamma}\sim 700\,$GeV, the GF counts about 6.4\% and 2.6\% of the LO results, for $\gamma\gamma$ and $\gamma\gamma j$ productions, respectively. The top quark mass in the GF calculations has been set to both 173\,GeV and infinity to show the mass effect, which can be seen more clearly in Fig.~\ref{rat}. The top quark effect is similar for both $gg\to\gamma\gamma$ and $gg\to\gamma\gamma g$ cases, which becomes significant near or above the threshold $M_{\gamma\gamma}\sim 2m_t$, and goes to the constant value when $M_{\gamma\gamma}\gg m_t$, as expected, i.e., $(\sum_1^3 q_d^2 + \sum_1^3 q_u^2)^2/(\sum_1^3 q_d^2 + \sum_1^2 q_u^2)^2=(15/11)^2=1.86$, with $q_u$ and $q_d$ represent the charge of the up-type and down-type quarks.

\begin{figure}[h]
\centering
\includegraphics[width=0.9\textwidth]{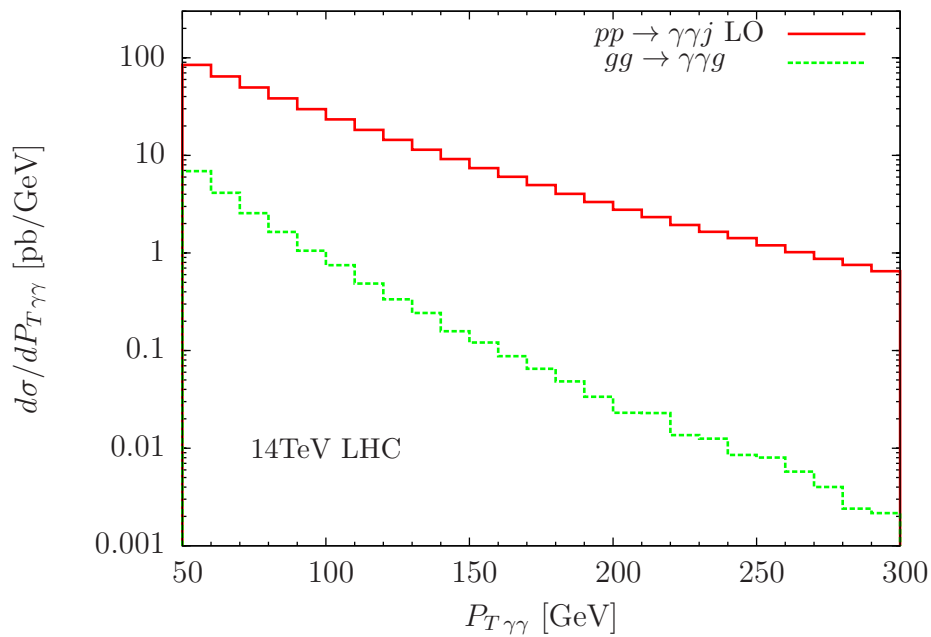}
\caption{$P_{T\,\gamma\gamma}$ distributions of the GF and LO QCD total results for $\gamma\gamma j$ productions at the 14 TeV LHC. }
\label{ptj}
\end{figure}

Fig.~\ref{ptj} shows the distributions of jet transverse momentum $P_{T\,\gamma\gamma}$, for $\gamma\gamma+j$ productions via GF at the 14 TeV LHC, in comparison with the LO QCD results. One can see that GF tends to have softer jet, due to the dilution effects of fermion loop. At $P_{T\,\gamma\gamma}\sim 50\,$GeV, GF counts 8.1\% of the LO result.

\section{Conclusions}
\label{summary}
We have checked and confirmed previous results on $gg\to \gamma\gamma$~\cite{dicus:1988}, $\gamma\gamma g$~\cite{deFlorian:1999tp} at one loop, using now the traditional Feynman diagram based approach, with special attentions paid to the numerical problem due to vanishing Gram determinants to get stable results. We updated those results for the 7 and 14 TeV LHC with new parton distribution functions and photo-jet isolation cuts. Our results show that the gluon-fusion process $gg\to\gamma\gamma g$ can contribute  $\sim 10\%$ of the Born result, especially at small $M_{\gamma\gamma}$ and $P^T_{\gamma\gamma}$, and increase further the overall scale uncertainty. Top quark loop effect is examined in detail, which has significant effects on the GF production rate itself when $M_{\gamma\gamma}\gsim 2m_t$. However, as the GF result drops quickly at large $M_{\gamma\gamma}$, the top quark mass effect on the GF+LO result is only at percent level.

\section*{Acknowledgments}
This work is partially supported by China Postdoctoral Science Foundation
under Grant No. 20100480466.

\section*{References}
\bibliographystyle{MyStyle}
\bibliography{biblio}

\end{document}